**Stability of 71° stripe domains in epitaxial BiFeO$_3$ films upon repeated electrical switching**

By *Florian Johann, Alessio Morelli*, and *Ionela Vrejoiu*

Florian Johann, Dr. Alessio Morelli
Max Planck Institute of Microstructure Physics
Weinberg 2, 06120 Halle (Germany)
E-mail: fjohann@mpi-halle.mpg.de

Dr. Ionela Vrejoiu
Max Planck Institute for Solid State Research
Heisenberg 1, 70569 Stuttgart (Germany) and
Max Planck Institute of Microstructure Physics
Weinberg 2, 06120 Halle (Germany)
Email: i.vrejoiu@fkf.mpg.de

*Abstract*
The 71° stripe domain patterns of epitaxial BiFeO$_3$ thin films are frequently being explored to achieve new functional properties, dissimilar from the BiFeO$_3$ bulk properties. We show that in-plane switching and out-of-plane switching of these domains behave very differently. In the in-plane configuration the domains are very stable, whereas in the out-of-plane configuration the domains change their size and patterns, depending on the applied switching voltage frequency. This paper has been published in Phys. Status Solidi B (http://dx.doi.org/10.1002/pssb.201248329)

Recently particular multiferroic domains patterns and multiferroic domain wall (DW) properties have moved into the focus of intense research. It has been observed that these domain patterns and DWs can have their own functional properties, dissimilar to the bulk domain properties.[1] In thin films the fraction of DWs compared to the film volume can become relatively high, and therefore the overall functional properties of the film may be dominated by the DW properties. Taking into account that DWs are in general mobile and can be manipulated for instance by an electric field, particular domain patterns may become promising for active memory devices.[2-3] As an example 109° and 71° stripe domain patterns of multiferroic BiFeO$_3$(BFO) have been investigated intensively, due to their unique properties. BFO has a rhombohedral structure with the direction of the ferroelectric polarization along the pseudocubic [111]$_C$ direction. As a result of the rhombohedral symmetry, the energetically favorable domain configurations in defect-free epitaxial BFO films consist of stripes, which are made of either 71° or 109° DWs.[4] In the last years there have been several reports about novel functionalities exhibited by 71° and 109° DWs. Seidel *et al.* demonstrated that native 109° DWs and 180° DWs are conductive, although the bulk of these domains is insulating. [5-6] Although the 71° DWs where found to be insulating by Seidel *et al.*, conductivity has been recently reported for 71° DWs of BFO films grown under different growth parameters.[7] Recently it has even been shown that 109° DWs exhibit a magnetoresistance effect.[8] Another peculiar property of a particular 71° stripe domain configuration in BFO films is related to an above band gap photovoltaic effect.[9] One of the main driving forces for the intensive research on multiferroics in the last decades is that they hold promise for magnetoelectric effects that are interesting for device applications. Heron *et al.* demonstrated that room temperature electrical switching of the polarization of 71° stripe domains of epitaxial BFO films can be used to reverse the magnetization of a CoFe layer in a CoFe/BFO heterostructure.[10] In this case, on one hand, not the value of the polarization of individual ferroelectric domains is relevant, but the yielded net in-plane value across the 71° stripe domain pattern is utilized. On the other hand, the switching of the polarization of the individual domains is coupled to the rotation of the antiferromagnetic order parameter of the BFO in the same and this is further coupled to the magnetization of the CoFe top layer, leading ultimately to the switching of the CoFe magnetization.

The reliability of switching of theses stripes is an important issue if BFO is considered to be used in device fabrication. However, not many studies have been reported so far regarding domain pattern stability. Shafer *et al*. reported on the possibility of studying the polarization switching between SrRuO$_3$ (SRO) bottom in-plane electrodes on a BFO/DyScO$_3$(110)$_O$ sample whose BFO film had 71° stripe domains.[11] Balke et al. reported on the in-plane switching of [110]$_C$-oriented BFO on SrTiO$_3$(110)$_C$, where switching between only two domain states was investigated.[12] Recently, Lee *et al*. investigated the in-plane switching of electrically leaky BFO films deposited on SrTiO$_3$(001)$_C$ and correlated the domains with local photoconductive measurements.[13] Folkman *et al.* investigated the electrical switching characteristics of 71° stripe domains and the effects of defect-dipoles on a BFO film grown on TbScO$_3$(110)$_O$, using interdigitated gold electrodes on top of the BFO film.[14]

However, there are no reports about the stability of these domains upon many electrical switching cycles and at different frequency of the applied voltage pulses. Here, we investigate the stability of 71° stripe domains in epitaxial thin BFO films for two configurations: after in-plane switching by applying an electrical field between in-plane electrodes aligned parallel to the 71° stripes (**Fig. 1b**) and after out-of-plane switching by applying the field perpendicular to the BFO film in plan-parallel capacitors (**Fig. 3a**). The latter configuration is preferable for applications in high density data storage devices. We show that the two electrical switching configurations have different behavior. For the in-plane configuration the 71° stripe domains can be switched repeatedly for many cycles at frequencies up to 100 kHz, resulting in only minor changes of the stripe domain pattern. For the out-of-plane configuration however, a marked dependence on the applied field frequency is observed. At frequencies below 1 kHz the 71° stripe domains are maintained, nevertheless the domain pattern of the film changes to stripes with larger widths compared to the as-grown state. For frequencies above 1 kHz the stripes can either be completely destroyed or the areas with same net-polarization break into smaller areas depending on the strain state and defect density.

For the in-plane configuration, 50 nm thick BFO films were grown directly on DyScO$_3$(110)$_O$ (DSO) substrate that were annealed in oxygen atmosphere before the deposition. The as-grown domain structure consists of 71° stripe domains along the orthorhombic [001]$_O$ direction of the substrate, with a net polarization pointing out-of-plane towards the film surface and in-plane along [$\bar{1}$10]$_O$ substrate direction (see schematics in Figs. 1a and 1b). The direction of the polarization could be revealed by vertical PFM (VPFM) and lateral PFM (LPFM) imaging (see Figs. 1d and 1e, respectively). It has been reported that annealing of the DSO substrates in O$_2$ atmosphere at elevated temperatures prior to depositing BFO can lead to formation of two ferroelastic domain variants with 71° stripe domain patterns in epitaxial BFO films.[9] However, the formation has not been studied much so far. It should be noted that the surface morphology (**Fig. 1c**) is very similar to BFO films with 109° stripe domains, which are obtained for BFO films grown on DSO substrates annealed in air (compare with Ref. 15). Here, for our BFO film with 71° stripe domains, the VPFM image in Fig. 1d reveals still a few small 109° stripes domains, which are aligned with the features in the morphology and perpendicular to the 71° stripe domains (short segments along $[\bar{1}10]_O$ with white contrast). Furthermore, for our BFO film with only 109° stripe domains on DSO substrate annealed in air, [15] the stripes are as well aligned along the $[\bar{1}10]_O$ axes, exactly as the few 109° stripe domains in the film with predominantly 71° stripe domains discussed here. From XRD measurements it can be deduced that for both samples the same structural variants are present (see Fig. S1 in the supporting information). This suggests that the DSO(110) substrate imposes in both cases the same structural/ferroelastic variants. If we consider one pseudo-cube of the DSO substrate, with its structure being monoclinic (Fig. 1a), the ferroelastic BFO domains are arranged in such a way as to adapt to the monoclinic distortion of the DSO(110). [15] The perpendicular orientation between the 109° stripe domains and the 71° stripe domains, which are built up from the same structural variants, is in agreement with the theoretical consideration by Streiffer et al. (see Figs. 2 and 4 in Ref. 4). Whether the 71° or 109° stripe domains are formed might be imposed by the DSO termination. It has been reported that the termination of the DSO substrate can be influenced by the atmosphere during annealing,[16] although other reports claim that for perfect single terminated DSO surfaces also a selective wet etching has to be applied.[17] Moreover, the resulting single-terminated $DyScO_3(110)$ surfaces are polar, being negatively charged if $ScO_2^-$ terminated, and positively charged if $DyO^+$ terminated.[17] In epitaxial rhombohedral ferroelectric films in general and in BFO films in particular, 71° and 109° stripes compete and the energetically more favorable configuration will form, depending on several parameters such as the film thickness and the electrostatic boundary conditions.[18-19] It is reasonable to assume that the different DSO terminations given by the different annealing procedures may affect the electrostatic energy terms, in particular the depolarization energy, and as a result stabilize different stripe domain patterns. Considering that the polarization of our 71° stripe domain BFO film on DSO substrate is pointing upwards, it might be inferred that our DSO substrate after $O_2$ annealing is mostly $DyO^+$ terminated.

The net in-plane polarization can be switched by applying a voltage to the in-plane electrodes. To investigate the domain development after switching, PFM was performed in an area between the electrodes on the as-grown BFO film and in the same area after repeated switching cycles. **Figure 2a** shows the LPFM image of the as-grown state and **Fig. 2b** the LPFM image after applying one unipolar rectangular pulse of +200V for 5 µs. The polarization of each stripe domain switched by 71°, resulting in a total 180° switching of the net in-plane polarization, as already reported.[10] After $5 \times 10^5$ switching cycles, which were performed by applying square pulses of ±200V at 100 kHz, the domain pattern hardly changed. Only at a few places, where the stripes were interrupted in the as-grown state of the BFO film, some small changes appeared, leading to partial removal of the initial interruptions, as can be seen in the comparison of the domain images in **Fig. 2c**.

The switching of the net in-plane polarization can be investigated by measuring also macroscopic in-plane switching currents. **Figure 2d** shows the switching currents collected for a hysteresis measurement performed at 1 kHz at room temperature after the BFO film was switched for $1.7 \times 10^6$ cycles. Assuming an effective polarization for this configuration of $P_{eff} = Q/(l*h)$, with $Q$ being the measured charge, $l$ the electrode length (1 mm) and $h$ the film thickness (50 nm),[14] values of ~50 µC/cm² for the polarization of BFO along $[\bar{1}10]_O$ are measured. The coercive voltages are very asymmetric and the resulting hysteresis loop for the polarization is thus strongly imprinted, indicating a built-in field aligned parallel to the direction of the as-grown net polarization. Imprint effects were attributed to defect dipoles for BFO films having 71° stripes domains, deposited on $TbScO_3(110)$.[14] **Figure 2e** shows the development of the coercive voltages and imprint with increasing cycle number. For the in-plane configuration, the imprint decreases after $1.7 \times 10^6$ cycles to ~66% of the original value. In addition, the coercive voltage for both polarities drops with increasing number of cycles as well. This improvement with switching cycles is an advantageous behavior with respect to device applications.

So far the conclusion is that in case of the in-plane switching configuration the domain pattern is very stable with improving electrical switching properties. However, in case of the out-of-plane electrical switching a steady transition of the as-grown stripe domain pattern to a new pattern occurs upon several switching cycles. For this configuration, 150 nm thick BFO was grown on $SrRuO_3$(SRO)-coated DSO(110)$_O$ and SRO-coated $SrTiO_3$(100)(STO). Copper top electrodes were thermally evaporated on top of the BFO films to form plan-parallel capacitors. We performed similar experiments with gold top electrodes and the results are summarized in Fig. S2 in the supplementary online material. For BFO films on both types of SRO-coated substrates, the as-grown domain structure consists of 71° stripe domains. The schematic in Fig. 3a describes the orientation of the 71° domains with respect to the DSO substrate and its surface terrace orientation. The VPFM images reveal that the polarization is pointing downwards to the SRO interface for both type of substrates (**Fig. 3b** and **4a**), as often reported for BFO films deposited on SRO-coated substrates.[15, 20-21] There are minor exceptions where some dark curved lines with polarization pointing upwards exist. These seem to have charged domain walls, as detailed in the supporting material (**Fig. S4**). These domains were seen by us before in BFO films grown on SRO/DSO and they seemed to be quite stable.[22] They appeared again after annealing the samples in vacuum at 350°C, subsequent to a chemical switching of the film polarization by oxygen plasma treatment. This points to a possible role played by oxygen vacancies in the stabilization of these domains. The LPFM image for the BFO film on SRO/DSO (**Fig. 3c**) shows that, unlike the 71° stripe domains of the BFO film grown directly on DSO, the 71° stripes are now aligned only along the $[\bar{1}10]_O$ direction, so they are 90° rotated. Additionally, there are regions with different net in-plane polarization directions, which are very often divided by the dark curved lines seen in the VPFM image. It can be deduced from LPFM images that at these lines the film often forms charged domain walls (see the domain analysis given in Fig. S4), however the PFM-measured domain pattern is too complex to allow an unambiguous assignment. TEM studies revealed that charged domain walls form in epitaxial BFO films, either in the as grown state[23] or after electrical switching.[24]

In the case of the BFO film on SRO/STO substrate there is predominately a single stripe direction with one net in-plane polarization direction. The direction of the stripes of BFO films grown on both substrates coated with SRO might be due to the epitaxial orientation and structure of the SRO layer, which is induced by the terraces of the substrate.[25-27]

Because lateral domain imaging was not possible through the top metal electrode, the electrode had to be etched away after the switching cycles. As a consequence, only the state after the cycling and removal of the electrode can be observed by PFM. The development of the domains in the same area of the BFO film after successive switching cycles cannot be directly observed. We had to investigate several electrodes that were used to switch the film with incremental number of cycles, subsequently etching the electrode away and then investigating the area underneath by PFM. The polarization switching of both Cu/BFO/SRO/DSO and Cu/BFO/SRO/STO substrates shows important frequency dependence. The BFO film on SRO/STO has much stronger frequency dependence than the BFO film on SRO/DSO. **Figure 3d** shows LPFM images taken in the areas where Cu/BFO/SRO/DSO capacitors have been cycled 5000 times at different cycle frequencies. At all frequencies there are still 71° boundaries existent, but the domain pattern has changed compared to the as-grown state. First, in the corresponding VPFM images (not shown here) a completely uniform contrast is seen with the out-of-plane polarization pointing to the SRO interface. The narrow dark lines belonging to domains of opposite out-of-plane polarization (Fig.3b) disappeared after the capacitor cycling. One reason for their disappearance might be that defects, e.g. oxygen vacancies, cation defects[13, 24] or cation impurities, which were produced during growth and were probably responsible for the formation of the curved narrow domains with different polarization in the first place, are mobile and redistribute during the switching cycles.[24] Secondly, for the lower frequencies of 0.1 kHz to 1 kHz, the areas with same net in-plane polarization grew laterally, whereas for higher switching frequencies these areas broke up into tiny areas.

**Figure 3e** shows the development of the stripe domains with the number of cycles at 0.1 kHz and **Fig. 3f** the average stripe width which has been extracted from these LPFM images. It can be seen that the average stripe width increased to the double of the as-grown value within the first 1500 to 2000 cycles and remained stable afterwards. However, the areas with same net in-plane polarization kept on increasing until the maximum performed switching cycles.

**Figure 4** summarizes the investigations performed on Cu/BFO/SRO/STO capacitors for which the frequency dependence of the stripe domains upon repeated switching was more dramatic. Figures 4a and 4b display VPFM and LPFM images of the as-grown state of the BFO film grown on SRO/STO. Figure 4c shows the LPFM images of Cu/BFO/SRO/STO capacitors that were cycled 5000 times at different frequencies. At 0.1 kHz the 71° stripe domain pattern was maintained. For the low frequency, the resulting stripe domain width also doubled, very similar to the Cu/BFO/SRO/DSO capacitors. In addition, for Cu/BFO/SRO/STO capacitors the direction of the net in-plane polarization was reversed after cycling, with respect to the as-grown state. From Fig. 4d it can be seen that for 0.1 kHz switching a similar number of pulses as for the DSO substrate was needed for the transition to the new domain state. Moreover, already for switching at 1 kHz the stripes began to break up, whereas at 10 kHz small disordered mosaic-like lateral domains formed.

**Figure 5** shows the electrical switching current measurements for the out-of-plane configuration performed on BFO films on both DSO and STO substrates. The voltage was applied to the top Cu electrode while the bottom SRO electrode was grounded. In the as-grown state (green curves in Fig. 5) the coercive fields were imprinted, indicating a built-in field pointing to the bottom electrode, parallel to the as-grown out-of-plane component of the net polarization. In addition, for negative voltages the leakage current increased with cycling. After $1.1 \times 10^4$ cycles, the coercive voltage increased at both switching frequencies (0.1 kHz and 10 kHz), but was more pronounced for the film on SRO/STO substrate. The magnitude of the leakage current improved for 0.1 kHz cycling, while for 10 kHz it worsened.

Figures 5b and 5d show the development of the coercive voltage and the imprint of BFO films grown on SRO-coated DSO and STO substrates, switched at 0.1 kHz and 10 kHz. It can be seen that, although the coercive voltages increased, the imprint stayed unchanged and their dependence on voltage pulse frequency was marginal.

It was established for other ferroelectric thin films, such as Pb(Zr,Ti)O$_3$ (PZT), that the alignment of defect-dipoles can strongly impact several material properties. For example, it can lead to enhanced voltage shifts, i.e. imprint. It was proposed that the net polarization determines the spatial location of the asymmetrically trapped charges that are the cause for the voltage shifts.[28]

The in-plane and the out-of-plane switching configurations show clear differences in the stability upon repeated switching at all frequencies. To understand the differences it is worth to have a closer look at the two configurations. The ratio between the electrode area, which should be proportional to the amount of nucleation centers, and the switched volume is more than two orders of magnitude higher for the out-of-plane electrode configuration. Therefore, in the out-of-plane configuration the BFO film will be switched in different areas of the electrode independently and then these areas have to merge together. If unfavorable domain configurations meet, the system needs to relax to a more favorable state via ferroelastic switching, which needs a certain relaxation time,[29] and thus leads to the stronger frequency dependence of the out-of-plane configuration.

Defects, especially the oxygen vacancies, may play a major role in the frequency dependence of the switching. Nelson *et al*. showed by *insitu* TEM investigations of switching of a BFO lamella by applying voltage between a needle and a planar bottom electrode that charged domain walls formed during the switching process and ordered planes of oxygen vacancies formed and acted as pinning centers.[24] Moreover, cation defects, such as Bi substitutions of Fe were detected in the vicinity of the 180° domain walls formed during the *in situ* TEM switching experiment. Lee *et al*. showed that the oxygen vacancies move and redistribute during in-plane switching of epitaxial BFO.[13] The effects of the oxygen vacancy motion and the formation of defect-dipoles in ferroelectric perovskites were extensively studied for other ferroelectrics, such as BaTiO$_3$ and PZT.[28,30] Oxygen vacancies and the defect-dipoles they may form were correlated with the degradation of the electrical resistance and also, indirectly, with the imprint of the ferroelectric capacitors, by affecting the potential wells for trapped electrons. It was also shown by Kim *et al.* that oxygen plasma exposure of as-grown BFO films led to switching the out-of-plane direction of the polarization.[22] A second step of annealing at 350°C in vacuum could recover the initial direction of the polarization, thus indicating the important role of oxygen content in the stabilization of preferential orientation of polarization in BFO films grown on $SrRuO_3$/$DyScO_3$(110). Additional evidence that defects play an important role is that another BFO film grown on SRO/DSO, which has more lines of opposite polarization in the as-grown state, i.e. most likely more defects, exhibited stronger frequency dependence and degradation of the stripe domains (see Fig. S3 in the supplementary information). The substrate on which the BFO film is deposited is likely to influence the switching behavior. Although for Cu/BFO/SRO/STO a similar amount of dark domains with opposite polarization exists in the as-grown state compared to the Cu/BFO/SRO/DSO sample, the frequency dependence is more dramatic for the former. The STO substrate subjects the BFO film to a higher compressive strain than the DSO substrate, due to the larger in-plane lattice mismatch. From X-ray diffraction reciprocal space mapping it was seen that the BFO film on SRO/STO is relaxed, whereas the BFO film on SRO/DSO is still almost completely coherently strained (see Fig. S1c and S1d in the supplementary information). The relaxation is usually accomplished by formation of misfit dislocations. We expect therefore a larger density of misfit dislocations to form in the BFO film grown on SRO/STO substrate, due to the larger in-plane mismatch.

Summarizing, we investigated the switching of 71° stripe domains in capacitor structures, both in in-plane configuration and in out-of-plane configuration. The two configurations show very different behavior for both the macroscopic electrical switching characteristics as well as the 71° stripe domain pattern stability. For the in-plane configuration the stripes are very stable for many switching cycles and independent of the investigated switching frequencies, up to 100 kHz. The coercive voltages and the imprint decreased with increasing number of switching cycles. For the out-of-plane configuration changes of the domain pattern occurred upon repeated switching within the first 2000-5000 switching cycles. For frequencies higher than 1 kHz the stripes were either completely destroyed or broke into smaller areas, depending on the strain state, defect type and defect density present in the film. For cycling at 0.1 kHz the 71° stripes were maintained, however the stripe width doubled with respect to the as-grown state. The coercive voltage increased with increasing number of cycles and the imprint of the hysteresis loops, however, stayed constant. The electrical resistance degradation improved for cycling at 0.1 kHz. The switching behavior and the limitations of the ferroelectric/ferroelastic $BiFeO_3$ domains unraveled here are very important in view of employing $BiFeO_3$ films with 71° stripe domain patterns in magneto-electric devices.[10]

*Experimental*

$SrTiO_3$(100) substrates were pretreated by etching in a buffered HF solution and annealing at 950°C for 2h. $DyScO_3$(110) substrates were prepared in two different ways: either annealed in $O_2$ atmosphere at 1250° for 3 hours and used for growing $BiFeO_3$ films for the in-plane switching configuration, or annealed in air at 1200°C for 2h and used for the out-of-plane switching configuration. All substrates were atomically flat after annealing. SRO and BFO were deposited by pulsed laser deposition (PLD) at 650°C in 0.14 mbar $O_2$ with 5 Hz laser repetition rate and a growth rate of 5 nm/min and 1.5 nm/min, respectively. After deposition the films were cooled to room temperature in 200 mbar $O_2$. Copper and gold top electrodes were deposited ex-situ by thermal evaporation through a shadow mask: for the in-plane configuration two 1 mm long electrodes were evaporated parallel to the stripe direction with a gap of 20 µm and for the out-of-plane configuration square shaped electrodes of size 50×50 µm². Switching currents were measured with an Aixxact TF2000 Analyzer. The voltage pulses for cycling the films were applied by a Tektronix AFG310 pulse generator, amplified if needed by the TF Analyzer amplifier. The rise time of the square pulses without amplifier were 75 ns, with amplifier about 200 ns. For the out-of-plane switching configuration the copper electrodes were etched with a diluted $(NH_4)_2S_2O_8$ solution and the gold electrodes were etched with gold etching solution from Alfa Aesar. PFM measurements were performed on a XE-100 Park scanning probe microscope in ambient conditions. Throughout the paper a bright (dark) contrast corresponds in VPFM to a polarization pointing downwards (upwards) and in LPFM to a polarization pointing to the right hand side (left hand side) of the image. XRD measurements were done with a Philips X'Pert diffractometer.

*Acknowledgements*
Financial support by the German Science Foundation in the framework of SFB 762 is gratefully acknowledged. We thank Dietrich Hesse for a careful reading of this manuscript.

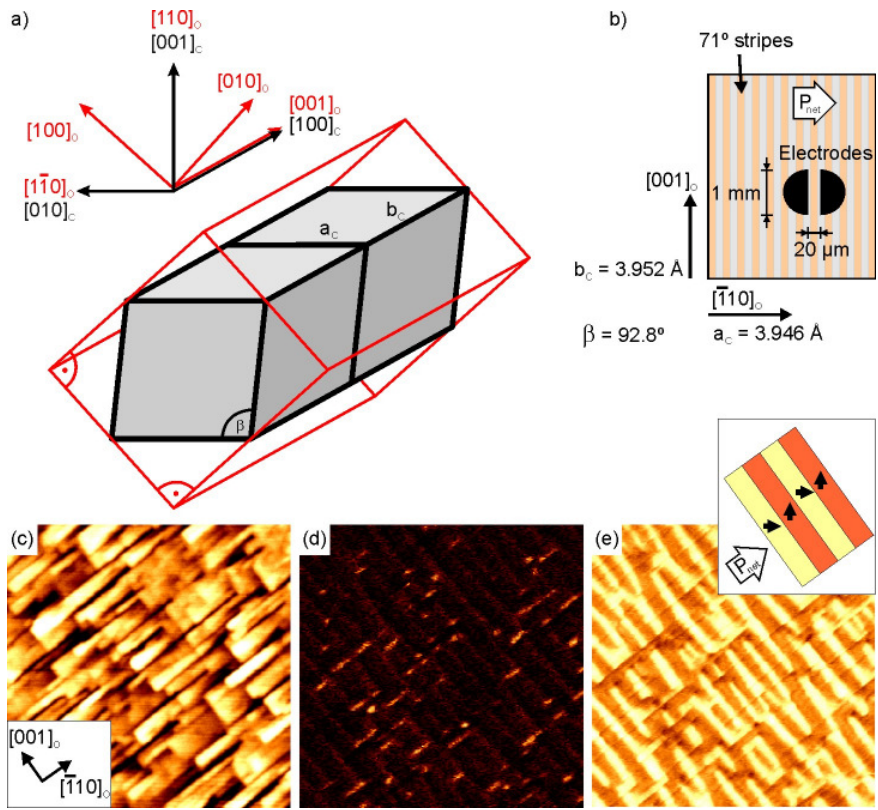

**Figure 1.** (a) Schematics of the orthorhombic $DyScO_3$ unit cell with two of the four monoclinic distorted perovskite cubes inside. (b) Top-view schematics of the $BiFeO_3/DyScO_3(110)$ sample with 71° stripe domain orientation and electrode alignment. (c) Topography, (d) VPFM phase and (e) LPFM signal images of the as-grown state of the $BiFeO_3/DyScO_3(110)$ sample. All images are 8×8 µm². Inset in (e) shows a scheme of the in-plane domain configuration.

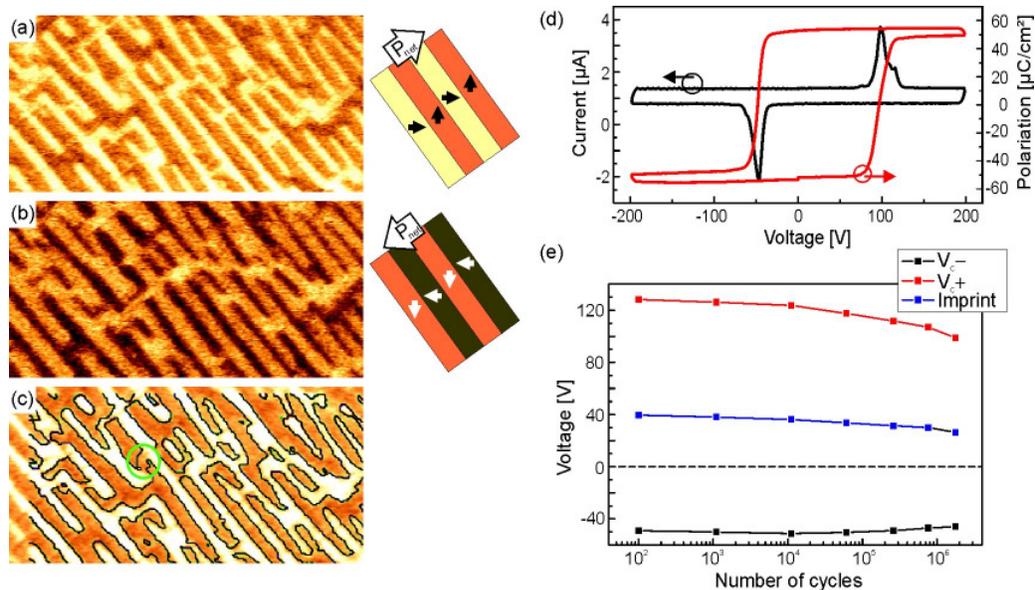

**Figure 2.** (a) LPFM image of as-grown state between the in-plane electrodes, (b) LPFM image after switching the net polarization with a unipolar voltage pulse and (c) LPFM image after $5\times10^5$ complete switching cycles with the domain walls of the as-grown state superimposed as black lines. The green circle indicates one of the few changes of the domain pattern. All images are 5×2.5 µm². (d) Switching current and integrated polarization of an in-plane hysteresis measured at 1 kHz and RT after $1.7\times10^6$ cycles on a second electrode. The capacitive charging contribution was subtracted from the polarization curve. (e) Development of the coercive voltages and imprint with number of cycles measured on the same electrode.

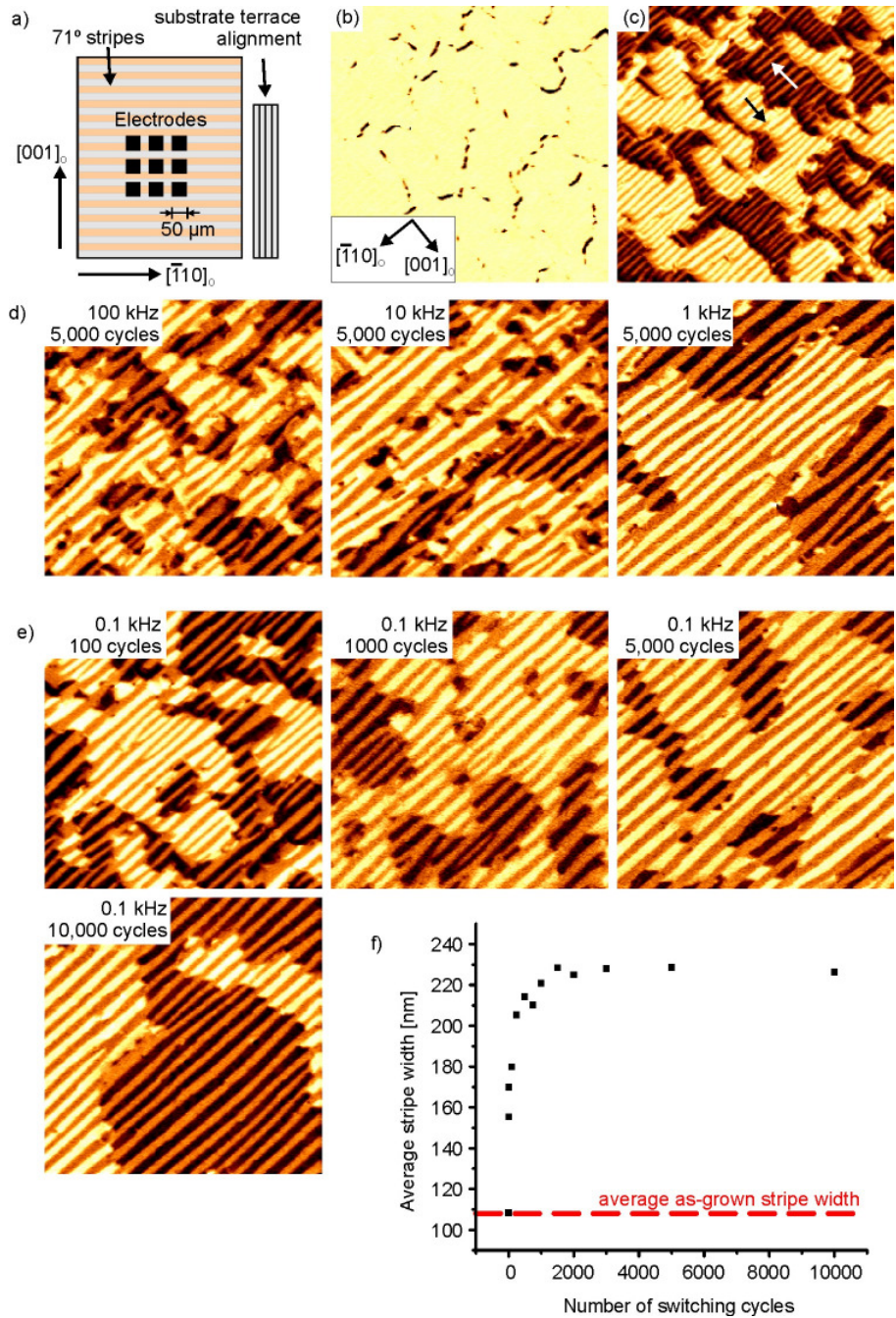

**Figure 3:** (a) Schematics of the Cu/BiFeO$_3$/SrRuO$_3$/DyScO$_3$(110) sample. (b) VPFM phase and (c) LPFM signal of as-grown state of the BiFeO$_3$/SrRuO$_3$/DyScO$_3$ sample. The arrows indicate the net in-plane polarization in the dark and bright regions. (d) LPFM images of capacitors which have been cycled 5,000 times at different frequencies and (e) capacitors cycled at 0.1 kHz with different number of cycles. All images are 8x8 µm². (f) Average stripe domain width as a function of the number of switching cycles at 0.1 kHz, measured for a BiFeO$_3$ film on SrRuO$_3$/DyScO$_3$(110).

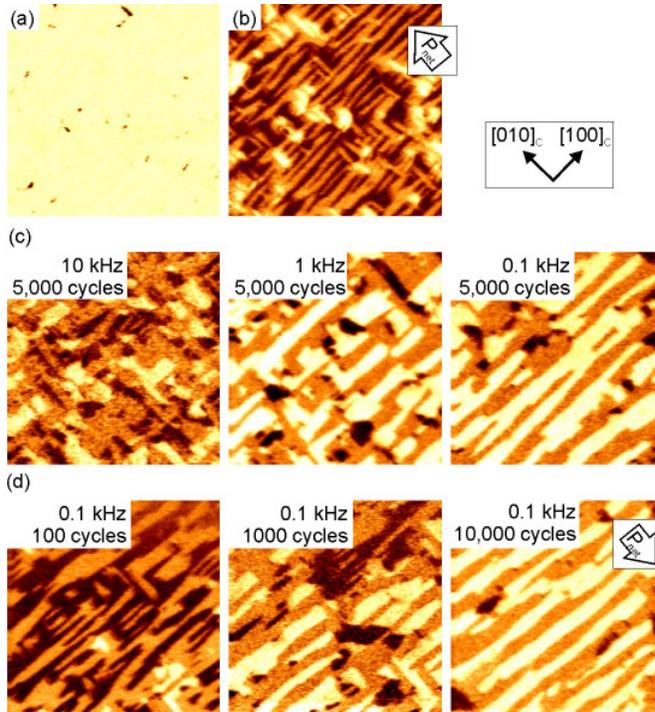

**Figure 4:** (a) VPFM phase and (b) LPFM signal of as-grown state of the BiFeO$_3$/SrRuO$_3$/SrTiO$_3$(001) sample. The arrow in (b) indicates the direction of the net in-plane polarization. (c) LPFM images of Cu/BiFeO$_3$/SrRuO$_3$/SrTiO$_3$ capacitors that have been cycled 5,000 times at different frequencies and (d) capacitors cycled at 0.1 kHz with different number of cycles. After 10,000 cycles the direction of net in-plane polarization is inverted compared to the as-grown state. All images are 6×6 µm².

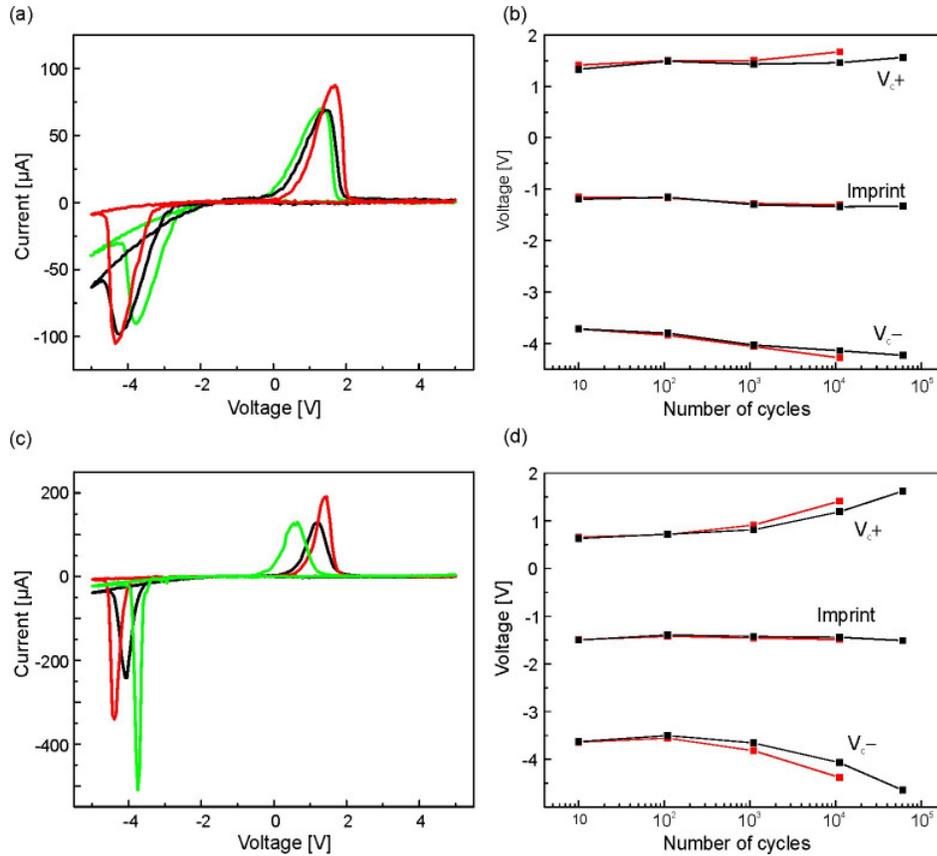

**Figure 5:** Measurement of the switching current at 1 kHz and room temperature for the sample (a) Cu/BiFeO$_3$/SrRuO$_3$/DyScO$_3$(110) and (c) Cu/BiFeO$_3$/SrRuO$_3$/SrTiO$_3$(100) with the as-grown hysteresis (green) and after $1.1 \times 10^4$ cycles with 0.1 kHz (red) and 10 kHz (black). The development of the positive coercive voltage V$_C$+ and the negative coercive voltage V$_C$− as well as the imprint is plotted for the film on (b) DyScO$_3$(110) substrate and (d) SrTiO$_3$(100) substrate.

Supporting Information:

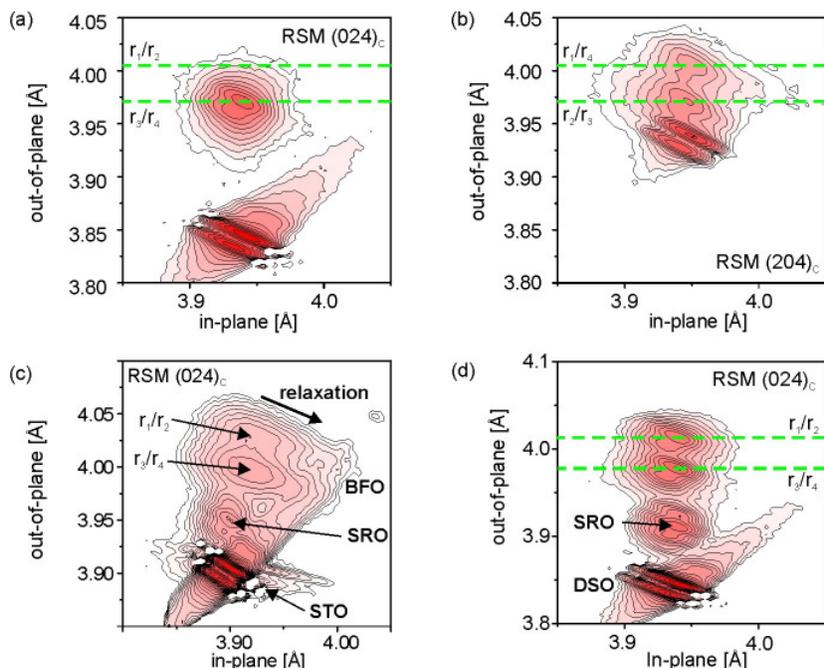

**Figure S1:** XRD reciprocal space maps (RSM) of the 50nm thick BFO film grown on $DyScO_3(110)_O$ substrate annealed in $O_2$ atmosphere around (a) peak $(024)_C$ and (b) peak $(204)_C$ for pseudocubic axis notation as defined in Ref 15. The same structural variants are present as for $BiFeO_3$ films grown on $DyScO_3(110)_O$ substrates annealed in air (compare with Fig. 5 in Ref. 15), suggesting that the structural variants are imposed by the monoclinicly distorted pseudocubic unit cell of the $DyScO_3$ substrate, irrespective of which stripes are formed. (c) RSM around $(024)_C$ of the 150nm thick $BiFeO_3/SrRuO_3/SrTiO_3(100)$ sample. Due to the preferential net in-plane polarization direction along the $[010]_C$ direction (See Fig. 4b in manuscript) the structural variants $r_1$ and $r_2$ are suppressed. In addition, the RSM shows the occurring relaxation of the $BiFeO_3$ layer, which is usually accomplished by forming misfit dislocations. (d) RSM around $(024)_C$ of the 150 nm thick$BiFeO_3/SrRuO_3/DyScO(110)_O$ sample. All structural variants occur, and the $BiFeO_3$ film does not show a relaxation of the in-plane parameter.

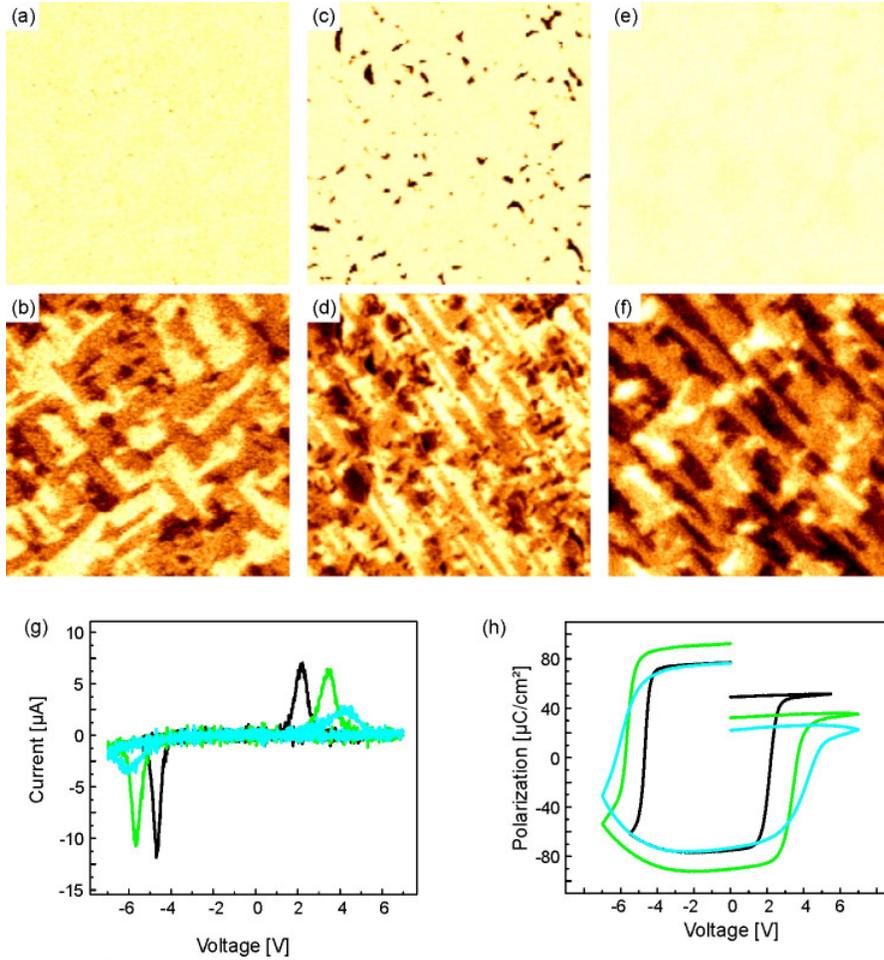

**Figure S2:** Comparison of Au and Cu top electrodes for the out-of-plane switching configuration: the 71° stripe domains are less stable in capacitors with Au top electrode than with Cu electrode. The VPFM image after 25,000 cycles at 10 kHz on (a) Cu/BiFeO$_3$/SrRuO$_3$/SrTiO$_3$ is still completely uniform and (c) for Au/BiFeO$_3$/SrRrO$_3$/SrTiO$_3$ capacitors opposite domains formed which can be seen as dark narrow domains of opposite polarization direction in the VPFM, similar to the as-grown state (see Fig. 4a in the manuscript). In-between, e.g. after (e) 5,000 cycles, the dark domains temporarily disappeared from the area. The transition of the lateral domains is similar for both electrode materials. (f) For Au top electrodes, after 5,000 cycles the stripes are already disordered as seen in the LPFM image, similar to Cu top electrodes. After the occurrence of the dark pinned domains, the lateral domains break up into even smaller domains for (d) compared to the (b) Cu top electrode. All images are 6×6 µm².

For Au/BFO/SRO/STO capacitors, with the appearance of the dark domains of upwards polarization, a decrease in switchable polarization is seen in macroscopic electrical switching measurements: in spite of the the leakage current contribution on negative polarity, (g) we measured smaller area under the switching current peak yielding a reduced (h) integrated polarization (black curve after 1,000 cycles, green curve after 5,000 cycles and blue curve after 25,000 cycles). A domain reconstruction of such lines seen in VPFM revealed that most of them have at least one charged domain wall (similarly also to the results in Fig. S4).

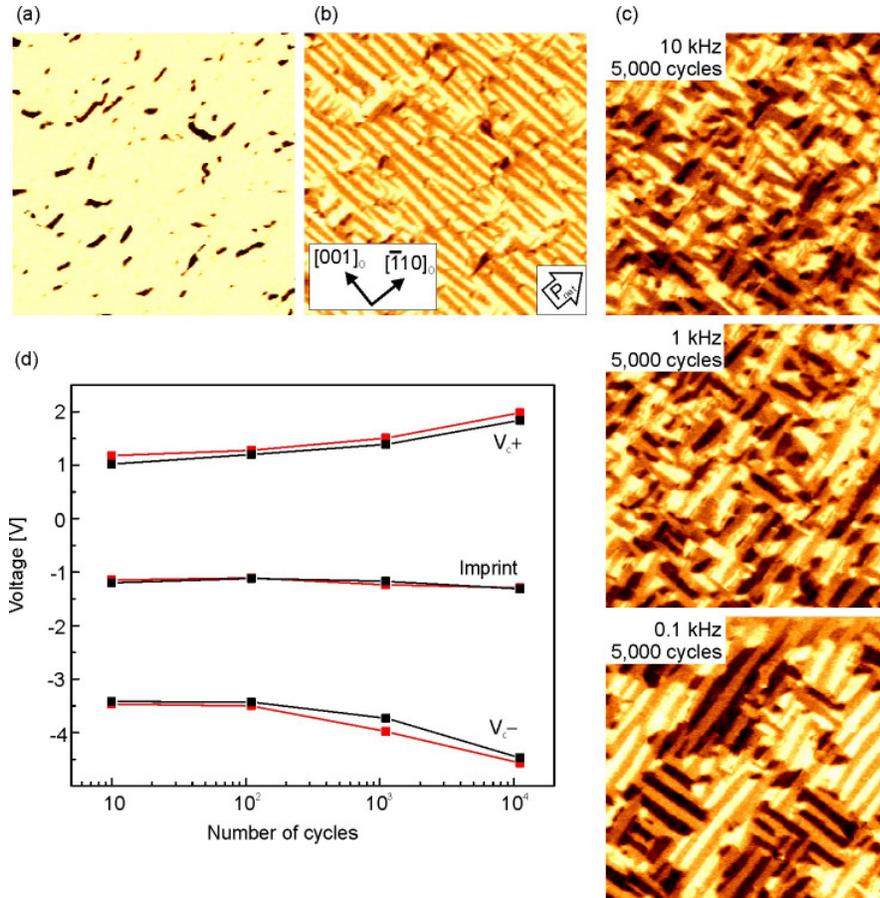

**Figure S3:** BiFeO$_3$/SrRuO$_3$/DyScO$_3$(110) film with more defects in the as-grown state compared with the film shown in Fig. 3, suggested by the lines of opposite polarization direction seen in (a) VPFM. LPFM images of (b) as-grown state and (c) after 5,000 cycles with different cycle frequency. All images are 8x8 µm². (d) Development of the coercive voltage and imprint with successive switching cycles on Cu/BiFeO$_3$/SrRuO$_3$/DyScO$_3$(110) capacitors. The frequency dependence is more pronounced than for the film with fewer defects.

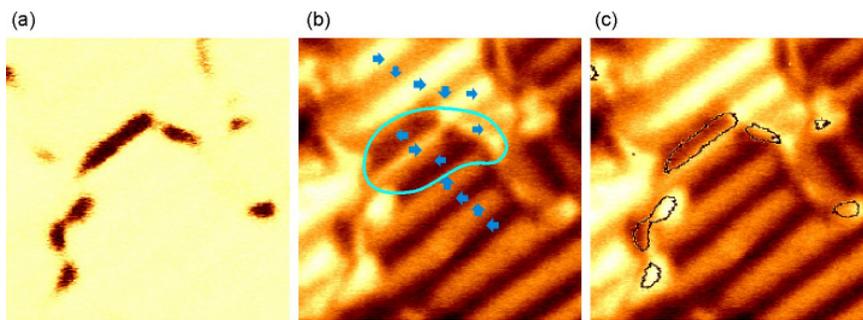

**Figure S4:** A zoomed-in PFM investigation (1.5 × 1.5 µm²) around a dark domain with polarization oriented upwards in the as grown state of BiFeO$_3$/SrRuO$_3$/DyScO$_3$(110) sample (see Fig. 3): (a) VPFM phase, where dark contrast indicates upwards out-of-plane polarization and (b) LPFM image with the in-plane direction of polarization indicated by arrows. It can be seen that many charged domain walls with head-to-head or tail-to-tail configurations exist in the vicinity of the dark domain seen in VPFM, marked by the light blue curve in the LPFM image. (c) Overlay of the location of the dark domains from the VPFM image onto the LPFM image.